\documentclass[aip, amsmath, amssymb, reprint,
]{revtex4-1}
\usepackage{float}
\usepackage{graphicx}% Include figure files
\usepackage{dcolumn}% Align table columns on decimal point
\usepackage{bm}% bold math
\usepackage[english]{babel}
\usepackage[utf8]{inputenc}
\usepackage[T1]{fontenc}
\usepackage{mathptmx}
\usepackage{etoolbox}

\makeatletter
\def\@email#1#2{%
 \endgroup
 \patchcmd{\titleblock@produce}
  {\frontmatter@RRAPformat}
  {\frontmatter@RRAPformat{\produce@RRAP{*#1\href{mailto:#2}{#2}}}\frontmatter@RRAPformat}
  {}{}
}%
\makeatother
\begin{document}

\preprint{}

\title[%Journal of Applied Physics
]{Portable Single-Beam Atomic Total-Field Magnetometer for Stand-off Magnetic Sensing}
% Force line breaks with \\
\author{Heonsik Lee}
\altaffiliation[~These authors contributed equally to this work]{}
\affiliation{OAQ Co. Ltd., Daejeon 34138, Republic of Korea
}

\author{Hyunbeen Lee}
\altaffiliation[~These authors contributed equally to this work]{}
\affiliation{OAQ Co. Ltd., Daejeon 34138, Republic of Korea
}

\author{Minseok Choi}
\affiliation{OAQ Co. Ltd., Daejeon 34138, Republic of Korea
}
\affiliation{Department of Physics, Korea Advanced Institute of Science and Technology, Daejeon 34141, Republic of Korea
}

\author{Yoontae Hwang}
\affiliation{Graduate School of Data Science,
Pusan National University,
Busan 46241, Republic of Korea
}

\author{Deok-Young Lee}
\email{dleeao@oaqcorp.com}
\affiliation{OAQ Co. Ltd., Daejeon 34138, Republic of Korea
}
\affiliation{Department of Physics, Korea Advanced Institute of Science and Technology, Daejeon 34141, Republic of Korea
}

%\date{\today}

\begin{abstract}
Optically pumped atomic magnetometers (OPAMs) offer high sensitivity at room temperature and are increasingly considered for portable magnetic sensing in geomagnetic-field environments.
Here we report a handheld-scale, single-beam scalar $^{87}$Rb OPAM with a sensor-head volume of approximately 110~mL.
The device operates in an all-optical Bell-Bloom configuration and uses digital lock-in, dispersive tracking of the $^{87}$Rb Larmor resonance, implemented with a hybrid electronics stack that combines in-house control hardware with commercial modules.
A single frequency-modulated laser beam performs both pumping and probing without RF coils.
All signal processing is realized in Python on a single-board computer paired with a commercial off-the-shelf (COTS) data-acquisition module, enabling immediate deployment without dedicated signal-processing hardware.
The magnetometer has an intrinsic in-band field sensitivity of approximately 21~pT/$\sqrt{\mathrm{Hz}}$, estimated from the lock-in dispersion slope, over a 0.1--30~Hz closed-loop in-band region with a digital-output rate of 200~samples/s.
In an unshielded Earth-field deployment, we detect repeatable transient magnetic signatures from a controlled elevator motion sequence and quantify standoff observability over sensor-elevator distances from 1.25~m to 10~m.
These results show that compact scalar OPAMs can provide bandwidth and range-resolved event sensitivity suitable for field-deployable magnetic anomaly detection and infrastructure monitoring in realistic geomagnetic environments.
\end{abstract}

\maketitle

\section{Introduction}

High-sensitivity magnetic-field measurement is central to applications spanning biomagnetism and magnetoencephalography (MEG) \cite{aslam_quantum_2023,sander-thommes_active_2023}, navigation and localization \cite{muradoglu_quantum-assured_2025,canciani_absolute_2016}, geomagnetic surveying and exploration geophysics \cite{kotowski_semi-airborne_2025,sharma_magnetic_1987, Stolz2022MinerEcon}, and indoor or infrastructure monitoring \cite{koss_optically_2022,huang_mains_2024}.
In many of these scenarios, the signal of interest is not the absolute field magnitude but a weak perturbation superimposed on the ambient geomagnetic background, induced by moving structures, electrical infrastructure, or nearby magnetic objects, and therefore must be detected in unshielded, time-varying environments \cite{oelsner_integrated_2022,rushton_unshielded_2022,Yao2022Opt}.
This magnetic-perturbation sensing regime emphasizes practical standoff observability, repeatability across trials, and robustness to drift and environmental variability, in addition to intrinsic sensor sensitivity.

Quantum-enabled magnetometers provide a pathway to improved precision relative to conventional magnetic sensors in low-signal settings \cite{degen_quantum_2017}.
Among them, optically pumped atomic magnetometers (OPAMs) based on alkali-vapor ensembles constitute a particularly mature and practical platform \cite{budker_optical_2007}.
OPAMs operate at room temperature and convert magnetic-field-induced atomic spin dynamics into an optical readout, enabling high sensitivity together with absolute calibration anchored in well-known atomic parameters \cite{dang_ultrahigh_2010}.
These attributes have motivated continued efforts toward compact, portable OPAM instruments suitable for geomagnetic-field operation and field-relevant measurements \cite{oelsner_integrated_2022}.

Earth-field scalar (total-field) OPAMs commonly employ synchronous optical pumping together with phase-sensitive (lock-in) detection \cite{yalaz_response_2024,schultze_optically_2017}.
In this approach, optical modulation drives a magnetic-resonance response, and the demodulated quadrature component provides a dispersive error signal that can be used to track resonance shifts in closed loop.
A foundational all-optical route is the Bell--Bloom magnetometer \cite{bell_optically_1961,yoon_laser_2025}, in which modulated light drives magnetic resonance without requiring an RF drive coil.
Recent work has further explored single-beam, all-optical pumping/probing strategies that use laser modulation to simplify the sensor head and reduce component count, while maintaining bandwidth suitable for transient sensing \cite{petrenko_single-beam_2021,zhang_zero_2022,wu_compact_2025, Kang2017JournalSensors, IEEE_TIM2019_singlebeam}.
Notably, dead-zone-free operation using single-beam free-induction-decay (FID) techniques has been demonstrated \cite{Mehta2025APL_deadzone}, and signal-enhancement schemes based on spin-polarization manipulation have been proposed to extend the applicability of single-beam FID magnetometers \cite{Jiang2025OL_FIDenhance}.

While these advances have significantly reduced the optical and mechanical complexity of OPAM sensor heads, the readout electronics of most portable implementations continue to rely on lock-in amplifiers or dedicated signal-processing hardware \cite{zanoni_picotesla_2024}, adding cost, volume, and expertise barriers that limit accessibility and field deployment.
A complementary strategy is to replace these specialized instruments with commercial off-the-shelf (COTS) data-acquisition modules and software-defined signal processing, accepting modest resolution trade-offs in exchange for immediate reproducibility and lower system complexity.

Distance dependence matters for two reasons: (i) magnetic perturbation amplitudes typically attenuate with increasing separation, and (ii) event separability depends on both the environmental background and the signal-processing pipeline that produces the reported magnetic-field time series \cite{budker_optical_2007,imajo_signal_2021}.
In low-SNR regimes, direct inspection of $\Delta B(t)$ can yield ambiguous onset/offset estimates, which degrades the reproducibility of event-resolved amplitude extraction and distance-scaling analysis.
To address this practical gap, model-agnostic timing markers can be constructed from the same measured time series to emphasize transient structure while preserving physical interpretability.
In particular, the short-time energy of the temporal derivative is sensitive to abrupt transitions, whereas windowed spectral-entropy measures quantify changes in spectral concentration during structured events.
These features, which have been widely adopted in the magnetic anomaly detection literature, can offer complementary perspectives depending on standoff distance and waveform morphology (e.g., sharply structured near-field signatures versus noise-contaminated far-field signatures), potentially aiding event localization when used alongside the raw magnetic-field time series.

In this work, we present a single-beam $^{87}$Rb scalar OPAM based on an all-optical Bell--Bloom scheme with digital lock-in dispersive tracking, implemented as a compact and modular instrument using a hybrid electronics stack combining in-house control electronics and commercial off-the-shelf modules.
We then report unshielded indoor measurements of elevator-motion-induced magnetic signatures as a representative infrastructure-scale magnetic-perturbation source.
Using event-resolved analysis supported by derivative-energy and spectral-entropy timing cues, we quantify standoff trends for multiple event classes and show that compact, localized door-related signatures approach dipole-like distance decay, while the larger elevator car/counterweight signatures exhibit distance-dependent effective slopes under door-referenced standoff definitions.
This study provides a practical benchmark for range-resolved magnetic-perturbation sensing in realistic indoor environments using a total-field OPAM, motivating compact atomic magnetometers for field-relevant magnetic anomaly detection and infrastructure event monitoring.

\section{Sensor Design and Characterization}

In the presence of a magnetic field, the atomic magnetic moment precesses about the field direction at the Larmor angular frequency $\omega_L = 2\pi f_L = \gamma B_0$, where $\gamma$ is the gyromagnetic ratio.
A scalar OPAM operates by measuring the Larmor frequency of an atomic ensemble.
Since $\gamma$ is well known for a given atomic species, the field magnitude $B_0$ can be recovered by a constant scaling.
For $^{87}$Rb, the Larmor frequency can be expressed as
\begin{equation}
f_L = \frac{\gamma}{2\pi}B_0,\qquad \frac{f_L}{B_0}=\frac{\gamma}{2\pi}\approx 7.0~\mathrm{kHz/\mu T}.
\end{equation}

To measure the Larmor frequency in a vapor cell, we employ an all-optical Bell--Bloom magnetometer together with a lock-in detection architecture.
In this configuration, a single elliptically polarized laser field provides both optical pumping and signal readout.
The magnetometer is driven by applying a periodic modulation at $\omega_{\mathrm{mod}}$ to the laser field.
Previous studies \cite{Grujic2013} have demonstrated Bell--Bloom operation using amplitude, frequency, or polarization (helicity) modulation; here, we adopt frequency modulation because it enables a compact and rapidly tunable driving scheme.
Accordingly, the laser frequency is modulated in the vicinity of the $^{87}$Rb D$_1$ resonance line at the modulation frequency $\omega_{\mathrm{mod}}$.

\subsection{Operating Principle: Single-Beam Optical Pumping}

An elliptically polarized beam can be regarded as a superposition of circularly and linearly polarized components.
The circularly polarized component optically pumps the atoms, orienting the magnetic moment either parallel or antiparallel to the laser propagation axis depending on the polarization helicity $\xi$.
For simplicity, we consider a circularly polarized beam with $\xi = +1$ propagating along $\hat{z}$ in the presence of a static magnetic field $\vec{B} = B_0 \hat{x}$.
The spin dynamics are then described by the Bloch equation \cite{Grujic2013}
\begin{equation}\label{BE}
\dot{\vec{S}} = \gamma \vec{S} \times B_0\hat{x} - \gamma_{r} \vec{S} + \gamma_p (t)\hat{z},
\end{equation}
where $\gamma_r$ is the spin-relaxation rate and $\gamma_p(t)= \gamma_1 \sin(\omega_{\mathrm{mod}}t)+\gamma_0$ is the time-dependent pumping rate induced by laser frequency modulation.

Gruji\'c and Weis \cite{Grujic2013} showed that Eq.~\eqref{BE} can be solved analytically in the low-power approximation as
\begin{equation}\label{Sz}
S_z(t) \approx S^{(0)}_z + S^{(A)}_z A(\omega_{\mathrm{mod}})\sin(\omega_{\mathrm{mod}}t)+ S^{(D)}_z D(\omega_{\mathrm{mod}})\cos(\omega_{\mathrm{mod}}t),
\end{equation}
for $\omega_{\mathrm{mod}} \approx \omega_{L}$, where
\begin{equation}\label{Ab}
A(\omega_{\mathrm{mod}})=\frac{\gamma_r}{(\omega_{\mathrm{mod}}-\omega_L)^2+\gamma _r^2}
\end{equation}
\begin{equation}\label{Dp}
D(\omega_{\mathrm{mod}})= \frac{(\omega_{\mathrm{mod}}-\omega_{L})}{(\omega_{\mathrm{mod}}-\omega_L)^2+\gamma _r^2},
\end{equation}
and $ S^{(0)}_z, S^{(A)}_z,$ and $S^{(D)}_z$ are constants that are independent of $\omega_{\mathrm{mod}}$.

\subsection{Lock-in Detection Architecture}

The linearly polarized component is used for signal readout.
For a spin-polarized vapor, the optical susceptibility experienced by the $\sigma^{\pm}$-polarized components of the probe field can be expressed as \cite{Grujic2013,Tsyganok2019}
\begin{equation}
\chi_{\pm} (t) = \chi^{(0)} \pm \chi^{(1)} S_z(t)+ \chi^{(2)}A_{zz}(t).
\end{equation}
Here, $\chi^{(0)}$, $\chi^{(1)}$, and $\chi^{(2)}$ are constants, while $S_z(t)$ and $A_{zz}(t)$ represent the vector and tensor polarizations of the atoms, respectively.
A linearly polarized field can be regarded as the superposition of two circularly polarized fields with opposite helicities, and the two components experience different refractive indices, given by $n = 1+\frac{\Re\{\chi\}}{2}$.
As a result, the polarization of light rotates as it propagates through a spin-polarized medium due to circular birefringence (Faraday rotation).
The rotation angle can be written as \cite{VanBaak1996}
\begin{equation}
\Delta \theta
= \frac{2\pi}{\lambda}\left(\frac{n_+ - n_-}{2}\right)L
= \frac{\pi}{2\lambda}\left(\frac{\Re\{\chi_{+}\}-\Re\{\chi_{-}\}}{2}\right)L,
\end{equation}
or
\begin{equation}\label{rot_arg}
\Delta \theta(t) = \frac{\pi L}{2\lambda}\chi^{(1)} \;S_z(t),
\end{equation}
where $L$ is the optical path length in the medium (i.e., the cell dimension) and $\lambda$ is the optical wavelength.

Equations~\eqref{Sz} and \eqref{rot_arg} show that the measured signal is sinusoidal, with information about the Larmor frequency encoded in its amplitude and phase.
Therefore, demodulation of the measured rotation signal is required to recover $B_0$.

The most conventional method for demodulating a sinusoidal signal is lock-in detection.
In lock-in detection, the measured signal is multiplied by a reference waveform at the same frequency, $S_{\mathrm{ref}}(t) = \cos (\omega_{\mathrm{mod}}t+\phi_{\mathrm{ref}})$, and then low-pass filtered to extract amplitude and phase information.

Depending on the choice of $\phi_{\mathrm{ref}}$, the demodulated output can be expressed as a linear combination of Eqs.~\eqref{Ab} and \eqref{Dp}.
Because our objective is to determine the Larmor frequency, we configure the lock-in detection to isolate the dispersive response in Eq.~\eqref{Dp} as a function of $\omega_{\mathrm{mod}}$.
We set $\phi_{\mathrm{ref}}=0$ and sweep $\omega_{\mathrm{mod}}$; the Larmor frequency is then obtained from the zero crossing of the resulting dispersive signal.
Under these conditions, the low-pass-filtered lock-in output is
\begin{align}
S_{\mathrm{disp}}
&=\mathrm{LPF}\{\Delta \theta(t)\, S_{\mathrm{ref}}(t)\}\nonumber\\
&= \theta_0\, D(\omega_{\mathrm{mod}})\, \mathrm{LPF}\!\left\{\frac{\cos\!\big(2\omega_{\mathrm{mod}}t\big)+1}{2}\right\}\nonumber\\
&= \frac{\theta_0}{2}\,D(\omega_{\mathrm{mod}}),
\end{align}
where $\theta_0$ is a proportionality constant that depends on optical parameters and the polarization response (cf.\ Eq.~\eqref{rot_arg}).

After the dispersion curve is acquired, the laser modulation frequency is held at the lock point, $\omega_{\mathrm{mod}}=\omega_L$, for subsequent magnetic-field measurements.
Because repeated acquisition of the full dispersion curve would substantially reduce the measurement bandwidth, magnetic-field variations are instead inferred from a linearized relation between the measured dispersive signal and the Larmor frequency in the vicinity of $\omega_{\mathrm{mod}}=\omega_L$.
When the magnetic field changes by $\Delta B$, the Larmor frequency shifts by
\begin{equation}
\Delta \omega_{L} = \gamma \Delta B.
\end{equation}
Near the lock point, the dispersive response is well approximated as linear:
\begin{equation}\label{lin_diff_1}
S_{\mathrm{disp}} \approx
\left.\frac{dS_{\mathrm{disp}}}{d\omega_{\mathrm{mod}}}\right|_{\omega_{\mathrm{mod}}=\omega_{L}}
\Delta \omega_{L}.
\end{equation}
This yields
\begin{equation}\label{lin_diff_2}
\Delta \omega_{L}
=
\frac{S_{\mathrm{meas}}}
{\left.\dfrac{dS_{\mathrm{disp}}}{d\omega_{\mathrm{mod}}}\right|_{\omega_{\mathrm{mod}}=\omega_{L}}},
\qquad
\Delta B = \frac{\Delta \omega_{L}}{\gamma}.
\end{equation}

This approach enables continuous tracking of magnetic-field variations without repeated frequency sweeps, thereby preserving bandwidth while providing stable, high-sensitivity readout.

\begin{figure}[t]
    \centering
    \includegraphics[width=1.0\linewidth]{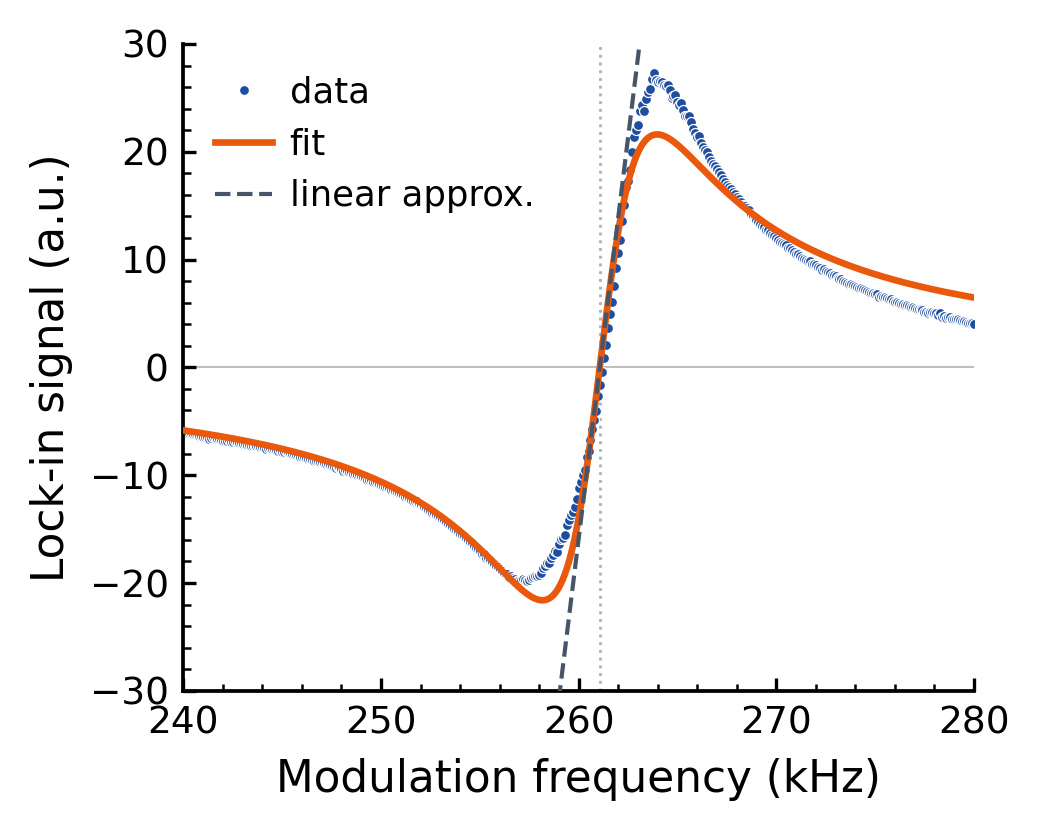}
    \caption{Lock-in-amplified Faraday rotation signal measured using the sensor as a function of laser modulation frequency, recorded inside a chamber (Twinleaf, MS-1LF) at $B_0 \approx 37.4~\mu$T. (Further details of the sensor implementation are provided in Subsection~\ref{sec:hw}.) Blue dots show the raw sensor readout. The orange curve is a fit to a dispersive Lorentzian, yielding a Larmor frequency of $f_L = 261.07~\mathrm{kHz}$ and a linewidth of $\mathrm{FWHM} = 5.81 \pm 0.08~\mathrm{kHz}$. The gray dashed line indicates a local linear approximation about $f_L$, with slope $|dV/df| \approx 14.9~\mathrm{a.u./kHz}$. The residual asymmetry of the dispersion feature suggests an additional background contribution consistent with a hole-burning (Bennett-structure) origin.}\label{fig:Disp}
\end{figure}

\begin{figure*}[t]
    \includegraphics[width = 1.0\linewidth]{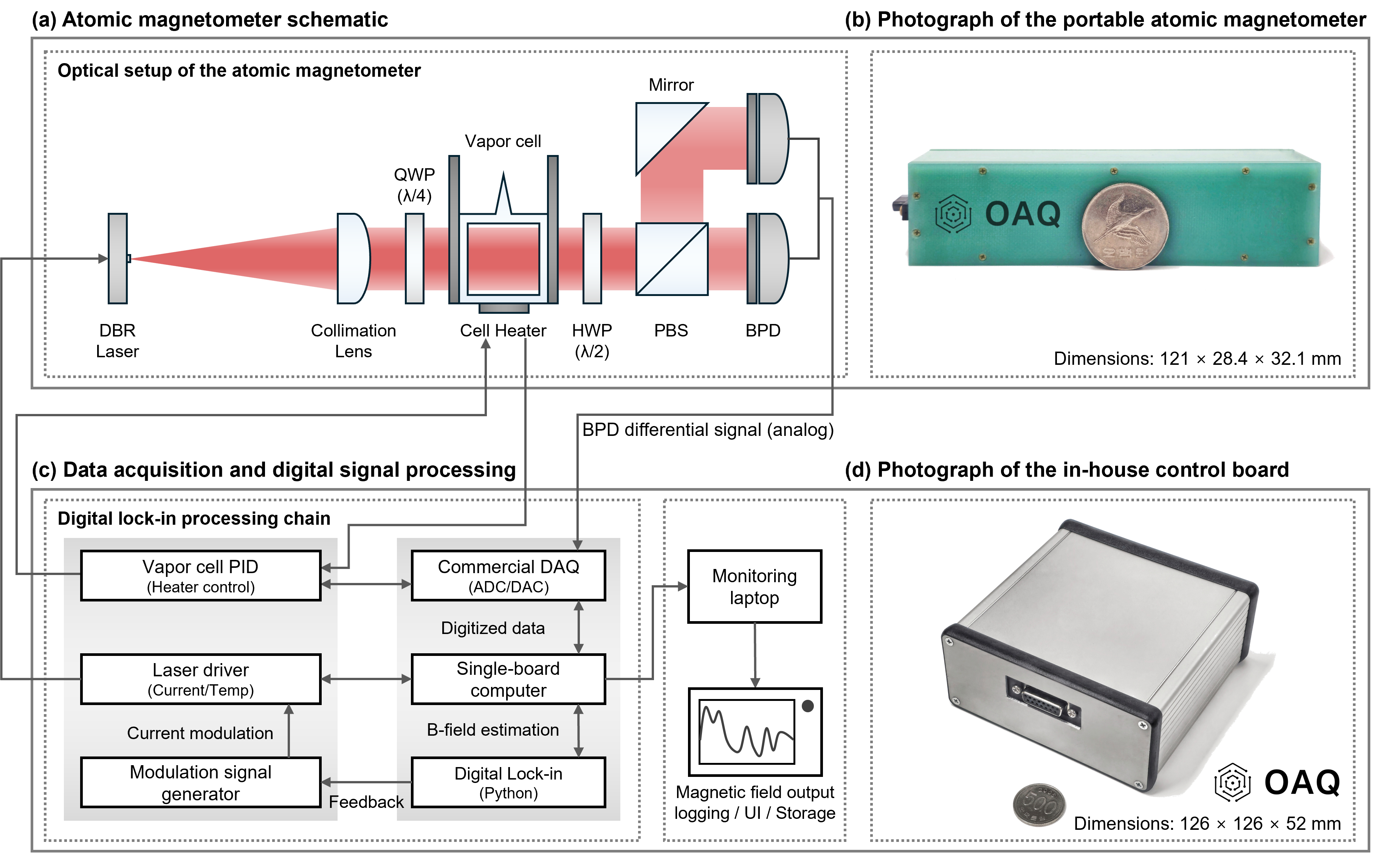}
    \caption{\label{fig:fig2}
    (a) Schematic of the portable single-beam atomic scalar magnetometer. The sensor head consists of a DBR laser, a collimation lens, a quarter-wave plate (QWP), an in-house fabricated $^{87}$Rb vapor cell with an integrated heater, a half-wave plate (HWP), a polarizing beam splitter (PBS), and a balanced photodetector (BPD).
    (b) Photograph of the assembled portable atomic magnetometer. The sensor has overall dimensions of 28.4 mm $\times$ 32.1 mm $\times$ 121 mm (approximately 110~ml).
    (c) Block diagram of the data acquisition and digital signal processing chain. The differential optical signal from the BPD is digitized by a commercial data acquisition (DAQ) device and processed on a single-board computer using Python-based digital lock-in detection to extract the magnetic-field signal.
    (d) Photograph of the in-house electronic control board used for laser current and temperature control, vapor-cell temperature regulation, and modulation signal generation. The board has overall dimensions of 126 mm $\times$ 126 mm $\times$ 52 mm and consumes approximately 5 W during sensor operation.}
\end{figure*}

Using the sensor described in Fig.~\ref{fig:fig2}, we measured the dispersion curve shown in Fig.~\ref{fig:Disp}.
The raw lock-in output is acquired without a DC offset, and the measured signal can be expressed as
\begin{equation}
S_{\mathrm{meas}}(\omega_{\mathrm{mod}})=S_{\mathrm{disp}}(\omega_{\mathrm{mod}})+\Delta S_{\mathrm{asym}}(\omega_{\mathrm{mod}}),
\end{equation}
where $\Delta S_{\mathrm{asym}}$ accounts for residual asymmetric contributions to an otherwise antisymmetric dispersive line shape $S_{\mathrm{disp}}$.
The asymmetry is commonly attributed to velocity-selective saturation (hole burning, i.e., Bennett-structure) effects \cite{BudkerYashchukZolotorev1998}, which can distort the line shape and shift the apparent zero-crossing point.
Such a shift introduces a potential bias in the inferred center frequency and thus limits the absolute accuracy of the magnetic-field readout.

For real-time magnetic-field sensing, however, we operate in a local linear regime by restricting the readout to a sufficiently narrow neighborhood of $\omega_{\mathrm{mod}}\approx\omega_{L}$, where the dispersive response is well approximated by its local slope, enabling differential tracking of small field-induced shifts.
In contrast, the Bennett-structure contribution manifests primarily as a slowly varying background \cite{BudkerYashchukZolotorev1998}.
Within a narrow window around $\omega_{\mathrm{mod}}=\omega_{L}$, it is therefore reasonable to treat the asymmetric background as locally constant to first order:
\begin{equation}
\left.\frac{d\Delta S_{\mathrm{asym}}}{d\omega_{\mathrm{mod}}}\right|_{\omega_{\mathrm{mod}}=\omega_{L}}\approx 0.
\end{equation}
Consequently, Eqs.~\eqref{lin_diff_1}--\eqref{lin_diff_2} remain valid provided that a compensation logic is included to correct for accumulated bias in $S_{\mathrm{meas}}$ (e.g., by tracking and subtracting slow drifts), and the resulting differential readout is sufficient for robust $\Delta B$ detection.

\subsection{Hardware implementation}\label{sec:hw}

Figure~\ref{fig:fig2}(a) shows the optical layout of the single-beam atomic magnetometer.
A distributed Bragg reflector (DBR) laser (Photodigm, 794.978 DBRHOT-L-CS) is used as a single optical beam for both optical pumping and probing.
The laser wavelength is tuned to the $^{87}$Rb D$_1$ transition near 794.98~nm, and the optical power at the vapor-cell entrance is approximately 10.5~mW.
To suppress magnetic-field disturbances originating from the laser package and associated drive electronics, the laser module is housed in a non-magnetic package \cite{Kim2023JAP}.
The laser output is collimated using an aspheric lens (Thorlabs, AC080-030-B), resulting in a beam diameter of approximately 5.3~mm.
Frequency modulation of the laser injection current enables single-beam operation.
The collimated beam is converted to circular polarization using a quarter-wave plate (QWP) before entering the atomic vapor cell.

The in-house-fabricated vapor cell \cite{Yoo2023copp, Lee_cubic, Yim2022AIP} contains isotopically enriched $^{87}$Rb and N$_2$ buffer gas at a pressure of 250~Torr.
The cell has a cubic geometry with a side length of 7.5~mm.
A resistive heater is mechanically attached to the vapor cell to maintain the operating temperature.
The heater geometry is designed to minimize and compensate the magnetic field generated by the heating current, reducing its impact on the magnetometer signal \cite{yim_note_2018}.
After transmission through the vapor cell, the optical beam passes through a half-wave plate (HWP) and a polarizing beam splitter (PBS), which separate the orthogonal polarization components.
The differential optical signal is detected using a balanced photodetector (BPD).
Balanced detection suppresses common-mode intensity noise and DC offsets, thereby improving the signal-to-noise ratio of the measured Faraday rotation signal.
The complete optical head is compactly mounted within a sensor head measuring 28.4~mm $\times$ 32.1~mm $\times$ 121~mm (approximately 110~mL).

The data acquisition and digital signal processing workflow is illustrated in Fig.~\ref{fig:fig2}(c).
The differential voltage output from the BPD is digitized by a commercial data acquisition (DAQ) device.
A multifunction DAQ board (Digilent, Analog Discovery 3) samples the signal at 15~MHz; digital lock-in processing uses an integration time of 250~$\mu$s to produce the magnetic-field time series reported below.
The acquired data are streamed to a single-board computer (SBC), where the magnetic field is extracted using Python-based digital signal processing.
The processing includes digital lock-in detection referenced to the applied modulation, enabling real-time demodulation of the dispersive magnetic resonance signal and conversion to magnetic-field values.

All sensor subsystems are controlled by an in-house electronic control board, shown in Fig.~\ref{fig:fig2}(d).
This board regulates the laser temperature, laser injection current, modulation signal, and vapor-cell temperature.
During operation, the laser temperature, injection current, and modulation amplitude are adjusted to tune the laser frequency to the $^{87}$Rb D$_1$ transition and to control the pump--probe interaction.
A sinusoidal modulation waveform is applied to the laser injection current to enable phase-sensitive lock-in detection, following standard modulation schemes used in OPAMs \cite{budker_optical_2007}.
The vapor-cell temperature is maintained at 93~$^\circ$C.
From a room-temperature start, the cell reaches its operating temperature within 2~min.

The complete control and readout electronics are fully self-contained and do not rely on rack-mounted laboratory instruments.
The in-house board has overall dimensions of 126~mm $\times$ 126~mm $\times$ 52~mm and consumes approximately 5~W during sensor operation.
This compact, low-power architecture enables portable operation while maintaining full digital control and real-time magnetic-field readout.

\begin{figure}[t]
    \centering
    \includegraphics[width=1.0\linewidth]{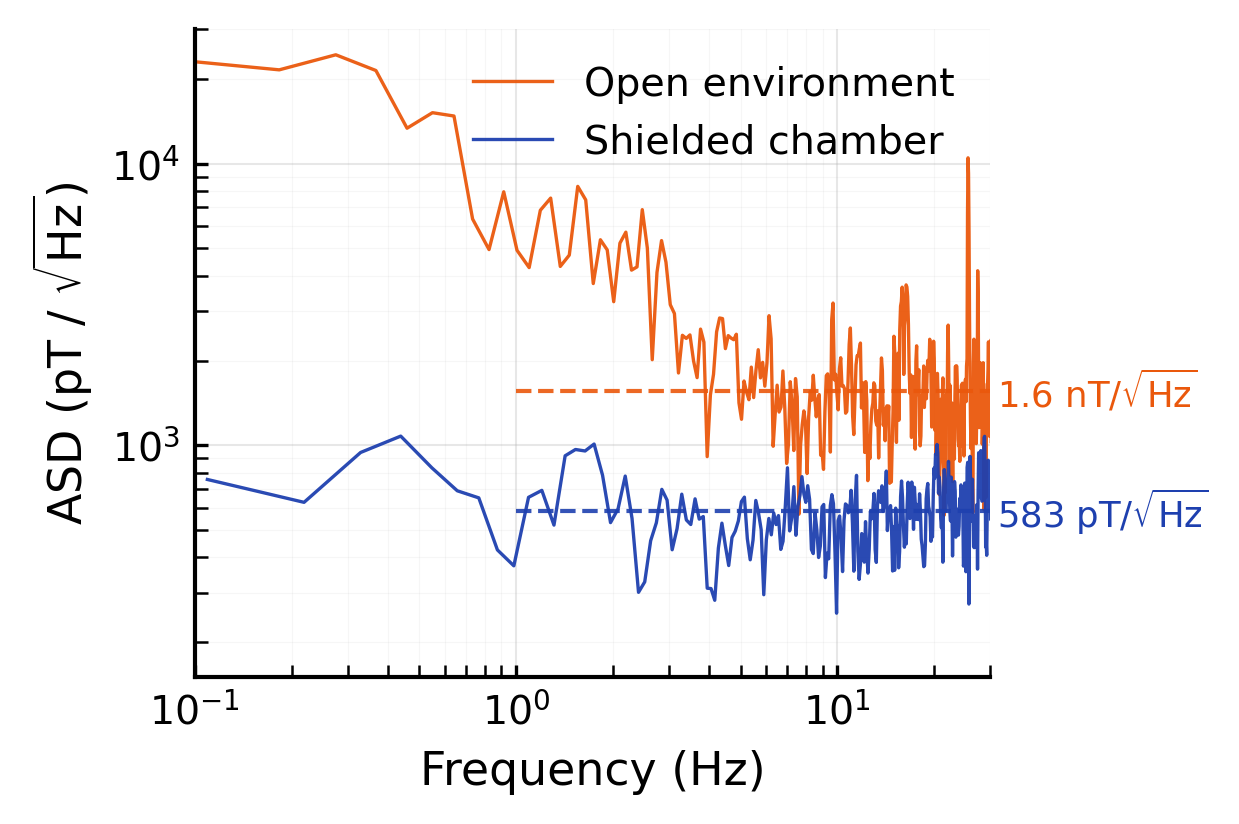}
    \caption{\label{fig:fig2-2a}
    Amplitude spectral density measured in the open laboratory environment (orange) and inside a chamber (Twinleaf, MS-1LF; blue), shown over the 0.1--30~Hz in-band region of the closed-loop magnetometer response. Dashed lines mark the directly measured in-band noise floors of approximately 1.6~nT/$\sqrt{\mathrm{Hz}}$ and 580~pT/$\sqrt{\mathrm{Hz}}$, respectively. Both directly measured floors are environment-limited; the intrinsic sensor sensitivity, estimated from the lock-in dispersion slope, is approximately 21~pT/$\sqrt{\mathrm{Hz}}$.}
\end{figure}

Figure~\ref{fig:fig2-2a} (orange trace) shows the amplitude spectral density (ASD) of the magnetic-field signal measured in an open laboratory environment.
The magnetometer is operated as a closed-loop digital lock-in system with a characteristic frequency $f_c \approx 30$~Hz set by the closed-loop tracking servo~\cite{budker_optical_2007}.
The trace is therefore reported within the 0.1--30~Hz in-band region, over which the closed-loop response remains approximately flat.
The in-band ASD reaches approximately 1.6~nT/$\sqrt{\mathrm{Hz}}$ and increases toward lower frequencies due to environmental drift.

To reduce these environmental contributions, an additional measurement was performed inside a chamber (Twinleaf, MS-1LF; Fig.~\ref{fig:fig2-2a}, blue trace) with $B_0 \approx 37.4~\mu$T.
In this configuration, low-frequency drift is partially suppressed and the in-band ASD becomes noticeably flatter, with a directly measured floor of approximately 580~pT/$\sqrt{\mathrm{Hz}}$.
This value is influenced by residual environmental and instrumental contributions (e.g., supply current noise, ferrite magnetization, and mechanical vibration), and thus represents the noise floor of the present measurement configuration.
This observation motivates the dispersion-slope-based estimate described below.

To estimate the intrinsic field sensitivity with reduced influence from the measurement environment, the lock-in dispersion slope at the operating point is employed~\cite{aleksandrov_modern_2009, schwindt_chip-scale_2007}.
For a single-shot capture of duration $T$ at the dispersion zero crossing, the equivalent field amplitude spectral density is given by
\begin{equation}\label{eq:disp_asd}
\mathrm{ASD}_B = \frac{\sigma_X \sqrt{2T}}{|dV/df|\, \gamma},
\end{equation}
where $\sigma_X$ is the standard deviation of the in-quadrature lock-in output and $\gamma$ is the gyromagnetic ratio.
Because $\sigma_X$ and $|dV/df|$ share identical instrumental units, the lock-in detection gain cancels exactly.
Applying this expression at the operating point ($T = 150~\mu$s) yields an intrinsic $\mathrm{ASD}_B \approx 21$~pT/$\sqrt{\mathrm{Hz}}$.
This value represents a lower-bound estimate rather than a directly measured free-running ASD.

\section{Results and Discussion}

To evaluate the performance of the portable atomic magnetometer in a realistic operating scenario, we investigated magnetic-field perturbations generated by the motion of an elevator in an unshielded building environment.
The measurements were designed to assess the capability of the sensor to resolve transient, event-related magnetic-field variations superimposed on the ambient geomagnetic background under real-world conditions.

\begin{figure}[!htbp]
\centering
\includegraphics[width=1.0\linewidth]{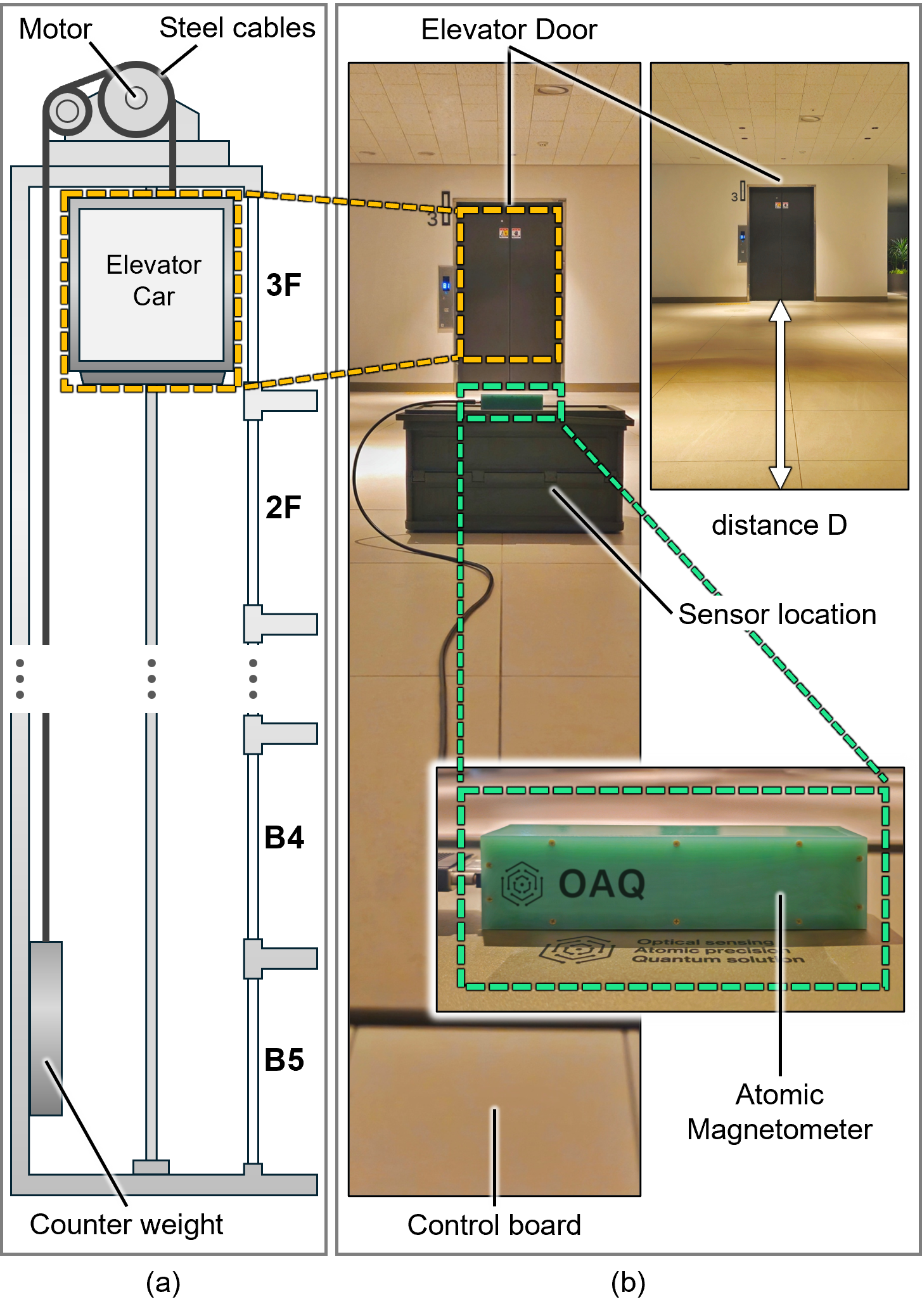}
\caption{\label{environment_photo}
Experimental setup for measuring elevator-induced magnetic-field disturbances using an atomic magnetometer.
(a) Schematic of the elevator shaft showing the elevator car, motor, steel cables, and counterweight, which constitute the primary sources of magnetic disturbances during elevator operation. The car and counterweight move in opposite directions along the shaft.
(b) Photograph of the measurement environment and sensor placement relative to the elevator door. The horizontal separation distance $D$ is defined as the distance from the elevator door to the magnetometer location. The sensor was mounted on a non-magnetic support structure and operated in an unshielded indoor environment.}
\end{figure}

The experiments were conducted in an eight-story building spanning from basement level 5 (B5F) to the third floor (3F).
The magnetometer was positioned on 3F near the uppermost section of the elevator shaft and in close proximity to the traction motor and drive components, as shown in Fig.~\ref{environment_photo}(a).
The elevator employs a machine-room-less (MRL) traction configuration with a counterweight (CWT) that moves opposite to the car.
Consequently, during a 3F$\rightarrow$B5F descent of the car, the CWT ascends along the shaft.
This coupled car--CWT motion produces distinct magnetic signatures that can be detected at 3F even when the car itself travels away from the sensor location.

Figure~\ref{environment_photo}(b) shows the experimental environment and the sensor placement.
Magnetic-field measurements were performed at multiple standoff distances $D$ between the sensor and the elevator door, ranging from 1.25~m to 7.5~m in increments of 1.25~m, with an additional measurement at 10~m.
The sensor head was mounted on a non-magnetic support structure to minimize magnetic coupling to the floor, and the signal was acquired in real time using an in-house electronic control board interfaced with a laptop computer.
In all measurements reported here, the laser beam axis was oriented approximately perpendicular to the ambient geomagnetic field direction, which maximizes the optical-pumping efficiency and the resulting magnetic-resonance signal amplitude.

It should be noted that the performance of a single-beam scalar magnetometer depends on the angle $\theta$ between the laser propagation axis and the ambient magnetic field.
When the beam is parallel to $\vec{B}$ (i.e., $\theta = 0^\circ$ or $180^\circ$), the transverse spin component that produces the precession signal vanishes, resulting in a ``dead zone'' characteristic of single-beam configurations \cite{Mehta2025APL_deadzone}.
In general, the signal amplitude scales approximately as $\sin\theta\cos\theta$ for orientation-based readout, so both the sensitivity and the measured perturbation amplitudes reported in this work correspond to the near-optimal orientation used.
Changing the sensor placement or rotating it relative to the geomagnetic field would alter the effective signal amplitude and could modify the observable distance range.
Quantifying the full orientation dependence and implementing dead-zone mitigation strategies (e.g., the FID-based approach demonstrated by Mehta~\textit{et~al.}~\cite{Mehta2025APL_deadzone}) are directions for future work.

\subsection{Event-resolved magnetic signatures in the time domain}

\begin{figure}[!htbp]
\centering
\includegraphics[width=1.0\linewidth]{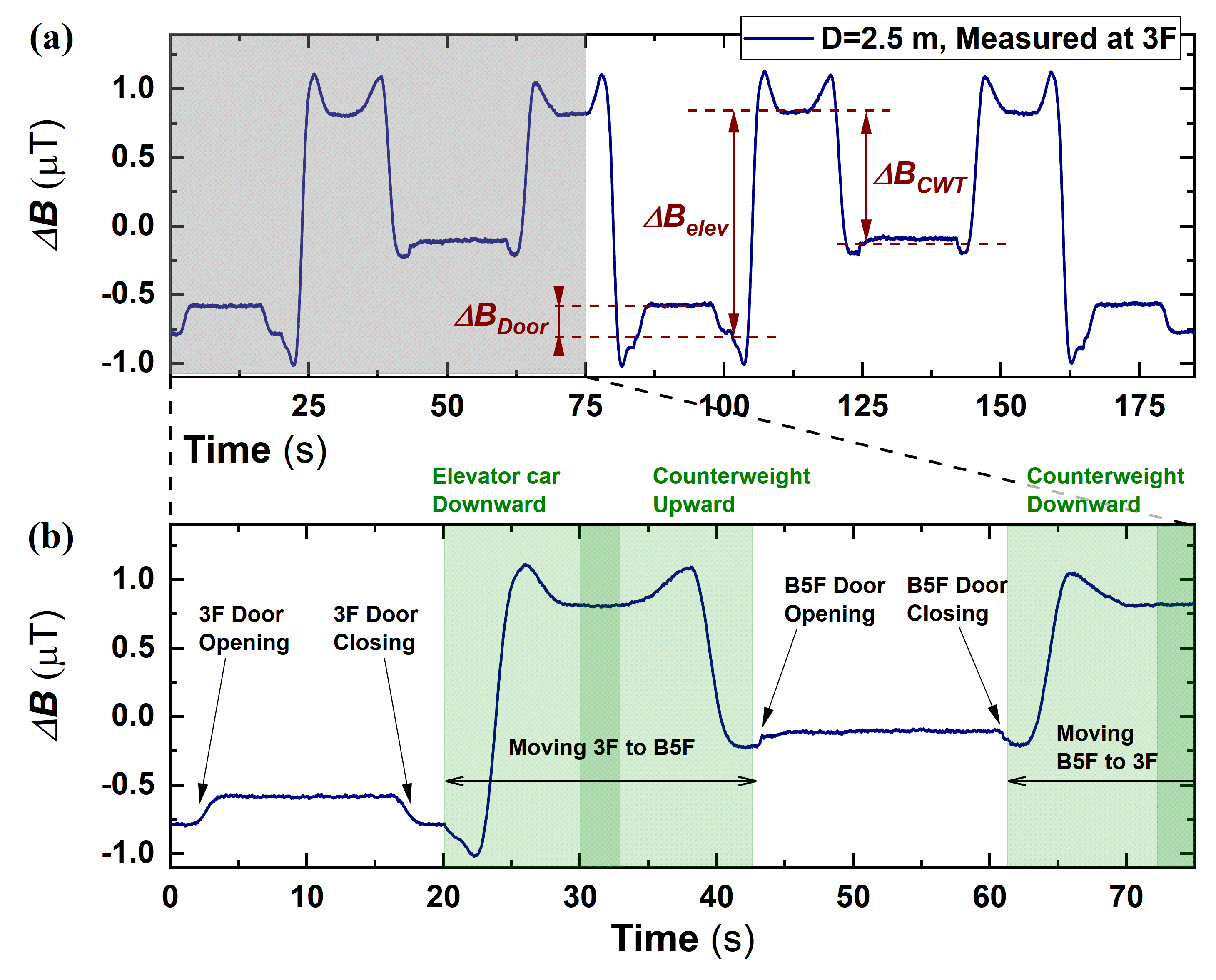}
\caption{\label{B(t)}
(a) Representative magnetic-field time trace ($\Delta B$) measured at a standoff distance of $D = 2.5$~m (sensor located on 3F) during an elevator operating cycle (3F$\rightarrow$B5F$\rightarrow$3F$\rightarrow$B5F$\rightarrow$3F). Step-like changes associated with 3F door operation ($\Delta B_{\mathrm{Door}}$), elevator car motion between 3F and B5F ($\Delta B_{\mathrm{elev}}$), and counterweight (CWT) motion ($\Delta B_{\mathrm{CWT}}$) are indicated.
(b) Expanded view of the 0--75~s segment highlighted in (a), capturing the sequence of 3F door opening/closing, car descent from 3F to B5F, B5F door opening/closing, and the onset of the return motion. Green shaded intervals correspond to the dominant mechanical states (car downward motion, CWT upward motion) used to identify the respective $\Delta B$ contributions.}
\end{figure}

Figure~\ref{B(t)} presents a representative time trace of the magnetic-field variation, $\Delta B(t)$, measured at $D=2.5$~m (sensor located on 3F) over one elevator operating cycle (3F$\rightarrow$B5F$\rightarrow$3F$\rightarrow$B5F$\rightarrow$3F), with the magnetic field recorded at 3F.
From this trace, three dominant classes of signatures are identified: localized door-operation events at 3F ($\Delta B_{\mathrm{Door}}$), motion-induced signatures associated with the elevator car during travel between 3F and B5F ($\Delta B_{\mathrm{elev}}$), and signatures arising from CWT motion ($\Delta B_{\mathrm{CWT}}$).
The expanded view in Fig.~\ref{B(t)}(b) highlights the temporal sequence of door operation, car descent, and coupled CWT ascent, enabling identification of the respective contributions.

When the elevator car is on the same floor as the sensor, opening and closing of the elevator doors produces distinct magnetic-field variations that are clearly resolved in the time domain.
At $D = 2.5$~m, the door-related magnetic-field changes reach amplitudes of approximately 0.2~$\mu$T.
As the standoff distance increases, the perturbation amplitude decreases, and door-operation signatures are no longer directly resolved for distances larger than approximately 5~m.
In contrast, magnetic-field changes associated with elevator motion remain observable at larger distances.
Representative time-domain data recorded at all standoff distances are provided in Appendix~\ref{sec:appendix}.

Following door operation, the elevator moves toward the lowest level (B5F).
During this motion, pronounced magnetic-field variations are observed across the measured distances.
The response reflects contributions from both the downward motion of the elevator car and the upward motion of the CWT, consistent with the MRL traction configuration, as indicated in Fig.~\ref{B(t)}.
At $D=2.5$~m, the peak amplitudes reach approximately 1.6~$\mu$T for the elevator car and 0.9~$\mu$T for the CWT, decreasing to approximately 0.08~$\mu$T and 0.045~$\mu$T, respectively, at $D=10$~m.
After the elevator reaches B5F, door-operation events at that level are only weakly detected at the sensor location on 3F.
When the elevator subsequently moves upward, the magnetic-field response exhibits a temporal profile that is approximately symmetric to that observed during downward motion, consistent with reversal of the elevator trajectory.

\subsection{Event timing markers using derivative energy and spectral entropy}

The raw $\Delta B(t)$ traces provide direct physical interpretability at short standoff distances; however, as $D$ increases, event-related perturbations approach the level of ambient fluctuations and manual identification of event onset/offset becomes less reproducible.
This motivates auxiliary timing markers that (i) do not rely on a detailed source model, (ii) remain sensitive to structured transient changes under low-SNR conditions, and (iii) enable consistent event-resolved amplitude extraction for distance-scaling analysis.
We therefore compute two complementary feature traces from $\Delta B(t)$: a derivative-based short-time energy $E(t)$ and a windowed spectral entropy $H(t)$ \cite{ZHOU2023}.

\begin{figure*}[t]
\centering
\includegraphics[width=1.0\linewidth]{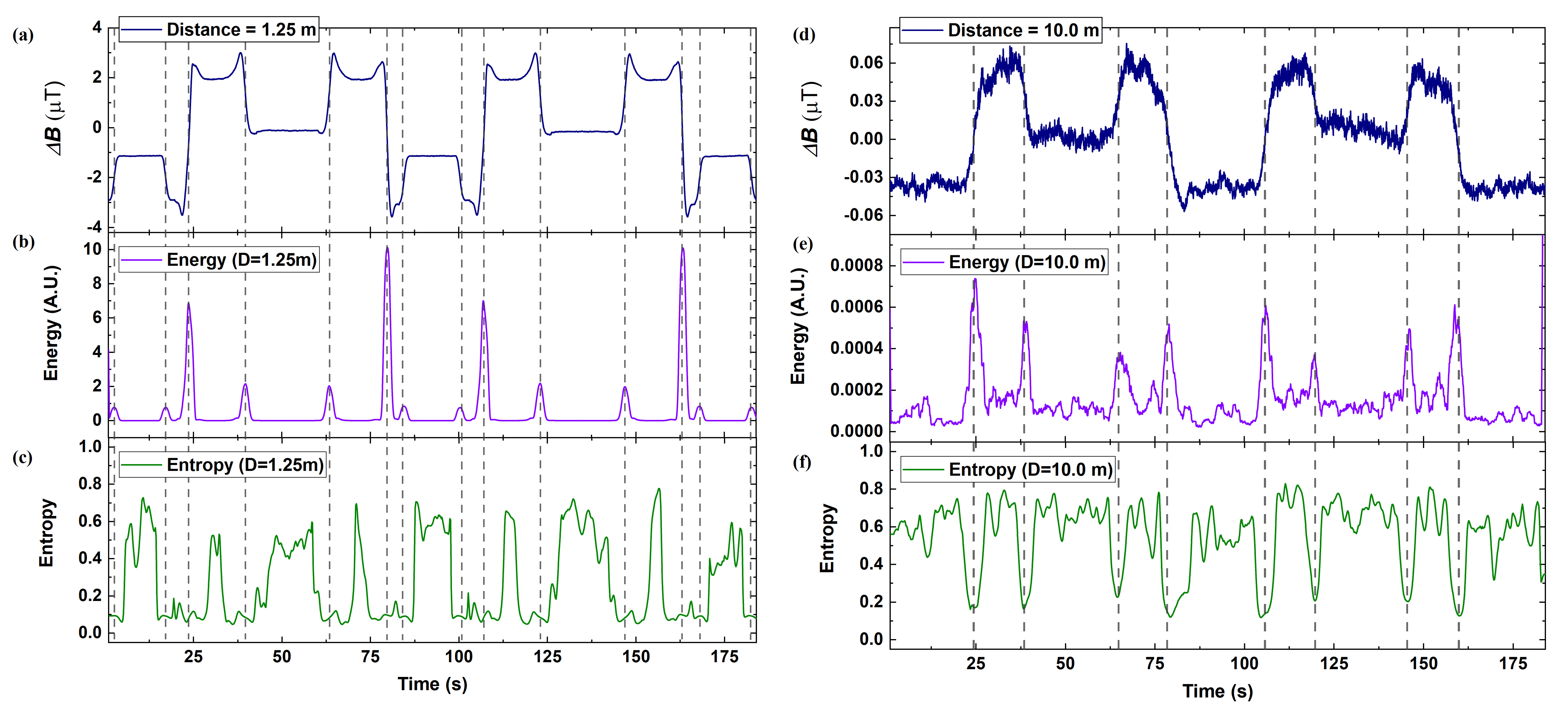}
\caption{\label{Height_comparison}
Comparison of elevator-induced magnetic signatures at two sensor--elevator distances. Panels (a)--(c) show results at $D = 1.25$~m, and panels (d)--(f) show results at $D = 10.0$~m. For each distance, the top row (a,d) presents the magnetic-field variation $\Delta B(t)$, the middle row (b,e) shows the short-time energy $E(t)$ computed from the temporal derivative of $\Delta B(t)$, and the bottom row (c,f) shows the normalized spectral entropy $H(t)$ computed from the windowed power spectrum. Vertical dashed lines indicate detected major events. At the shorter distance (1.25~m), the derivative-energy trace exhibits sharper and higher peaks at transition instants, whereas the entropy trace tends to remain near low values. At the longer distance (10.0~m), both energy peaks and entropy modulation remain visible. The entropy trace provides complementary information by reflecting changes in spectral concentration, although the raw $\Delta B(t)$ signal may remain more directly interpretable depending on the specific event and noise conditions.}
\end{figure*}

The derivative energy is designed to emphasize transition instants by converting step-like or rapidly varying segments of $\Delta B(t)$ into sharp peaks.
We compute
\begin{equation}
E(t)=\left\langle \left(\frac{d\Delta B}{dt}\right)^{2}\right\rangle_{T_E},
\end{equation}
i.e., a sliding-window average of the squared time derivative over a window duration $T_E$.

In practice, $\Delta B(t)$ is lightly smoothed to suppress high-frequency noise prior to numerical differentiation, and we use $T_E=2$~s.
Conceptually, $E(t)$ functions as a simple energy detector for abrupt changes, analogous to energy-based detectors widely used in magnetic anomaly detection and related transient detection problems \cite{rs16020363}.

In contrast, spectral entropy quantifies changes in spectral structure.
We compute the Shannon entropy of the normalized power spectrum within a sliding window and normalize it to the range $[0,1]$:
\begin{align}
H(t) &= -\frac{1}{\log M}\sum_{k=1}^{M} p_k(t)\log p_k(t), \\
p_k(t) &= \frac{P_k(t)}{\sum_{j=1}^{M} P_j(t)},
\end{align}
where $P_k(t)$ is the windowed power spectral density and $M$ is the number of frequency bins.
Spectral entropy has been used as an irregularity/structure measure in time-series analysis by interpreting the PSD as a probability mass function and applying Shannon entropy \cite{Wang2021AIP, Fan2020JoS}.
In this work, $H(t)$ is evaluated using a 4~s Hann window with a 0.2~s hop size and then interpolated to form a continuous time trace.

Figure~\ref{Height_comparison} compares $\Delta B(t)$, $E(t)$, and $H(t)$ for two representative cases ($D=1.25$~m and $D=10.0$~m), illustrating how these features complement each other across distance.
At $D=1.25$~m, the elevator-induced perturbation is strongly structured and dominated by step-like components; consequently, $E(t)$ yields high-contrast peaks at transition instants, while $H(t)$ often saturates near low values because the spectrum becomes highly concentrated (i.e., less entropic).
At $D=10.0$~m, the perturbation amplitude is reduced and more affected by ambient fluctuations, yet the transition peaks in $E(t)$ remain detectable.
In some intervals, $H(t)$ reflects changes in spectral concentration that complement the direct time-domain signal; however, it should be noted that the raw $\Delta B(t)$ trace itself remains informative and, depending on the specific event morphology and noise conditions, may provide comparable or even more direct identification of event boundaries.
Derivative energy and spectral entropy are therefore best understood not as replacements for the raw signal but as auxiliary timing features drawn from the broader magnetic anomaly detection literature \cite{ZHOU2023, Wang2021AIP, Fan2020JoS}, whose relative utility depends on standoff distance, event type, and ambient noise characteristics.
Thus, $E(t)$ primarily serves as a precise marker of rapid transitions, whereas $H(t)$ provides a complementary spectral-domain perspective that can support event identification under specific low-SNR conditions.

\subsection{Distance scaling of event amplitudes}

\begin{figure}[t]
\centering
\includegraphics[width=1.0\linewidth]{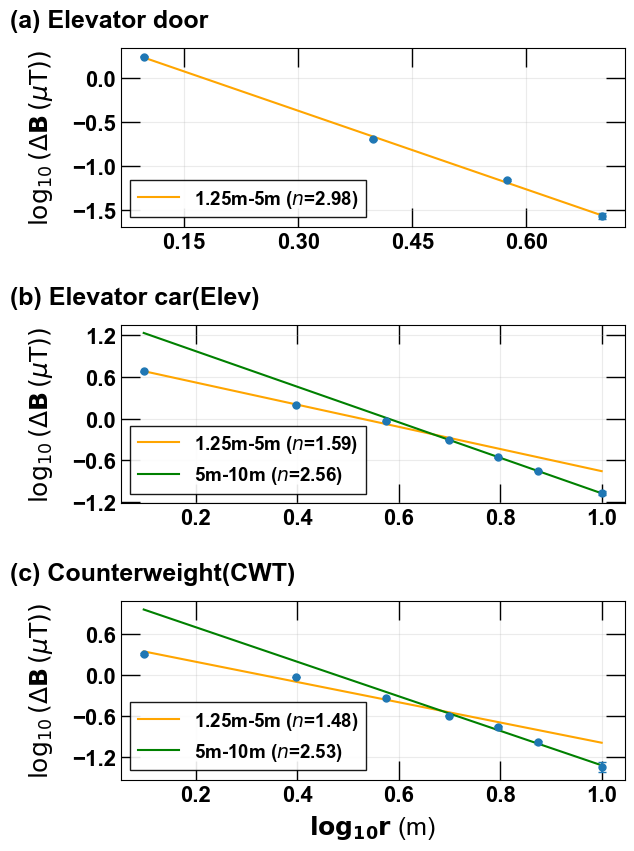}
\caption{\label{fig:fig3-3}
Distance dependence of magnetic-field perturbation amplitudes extracted from event-resolved measurements associated with (a) elevator door opening and closing, (b) elevator car motion, and (c) counterweight motion. Blue markers represent measured data, with error bars indicating measurement uncertainty. Solid lines indicate linear fits performed over different standoff distance ranges, as labeled in each panel. The door-related signal exhibits a clear $D^{-3}$ scaling over the measured distances, consistent with a localized dipole-like magnetic source in the far field. In contrast, signals associated with the elevator car and the counterweight show weaker distance dependence at shorter distances, while approaching an effective scaling exponent of $n \approx 2.5$ for standoff distances larger than approximately 5~m. This behavior is attributed to the extended geometry of these components and to the difference between the door-referenced standoff distance and the effective distance to distributed magnetic sources.}
\end{figure}

Using the event taxonomy identified from the magnetic-field time traces---namely, elevator door operation, elevator car motion, and CWT motion---together with the timing cues provided by $E(t)$ and $H(t)$, we extracted the magnetic-field perturbation amplitudes for each event class at each standoff distance ($\Delta B_{\mathrm{Door}}, \Delta B_{\mathrm{elev}}, \Delta B_{\mathrm{CWT}}$).
Figure~\ref{fig:fig3-3} summarizes the resulting distance dependence on logarithmic scales.

For the 3F door events, the extracted amplitudes exhibit a clear dipole-like decay with distance, approaching $\Delta B_{\mathrm{Door}} \propto D^{-3}$ over the measured range.
This behavior is consistent with the far-field scaling expected for an effectively compact magnetic disturbance near the entrance region \cite{Martinez2020ISMRM}.

In contrast, the elevator car and CWT signatures do not follow a single power law over the full distance range.
The log--log plots show a distance-dependent effective slope: the piecewise fit yields a shallower trend for $D \leq 5$~m ($n_{\mathrm{eff}}\approx 1.5$) and a steeper trend for $D > 5$~m (increasing toward $n_{\mathrm{eff}}\approx 2.5$).
Accordingly, the distance dependence in Fig.~\ref{fig:fig3-3} is interpreted in terms of apparent decay behavior rather than a single universal exponent.

The dipole far-field approximation is generally valid only when the sensor--target separation is sufficiently larger than the characteristic dimension of the target.
For extended ferromagnetic structures, near-field and distributed-source effects can lead to non-dipolar decay over practical standoff ranges, motivating numerical or calibrated modeling approaches in magnetic anomaly detection studies \cite{CHEN2022113806, s24124028}.
In this context, our in situ, event-resolved amplitude benchmarks and distance-robust timing markers provide a practical basis for validating and calibrating model-based distance predictions in realistic indoor environments.

\section{Conclusion}\label{sec:conc}

We developed and demonstrated a compact single-beam $^{87}$Rb scalar optically pumped atomic magnetometer (OPAM) based on an all-optical Bell--Bloom scheme with digital lock-in dispersive tracking.
The instrument integrates a simplified single-beam optical head with an in-house control board and a fully software-defined signal chain implemented in Python on a COTS data-acquisition module, enabling portable operation in the Earth-field regime while maintaining stable, real-time readout of magnetic-field variations.

To evaluate field-relevant performance in an unshielded and time-varying indoor environment, we measured elevator-operation-induced magnetic perturbations in an eight-story building with the sensor positioned on the third floor near the upper shaft region.
Across standoff distances from 1.25~m to 10~m, the raw time traces revealed repeatable, event-resolved signatures attributable to door operation, elevator car motion, and counterweight motion in an MRL traction configuration.
To support event timing analysis under reduced SNR at larger distances, we employed two lightweight, model-agnostic feature traces drawn from the magnetic anomaly detection literature: a derivative-based short-time energy $E(t)$ that emphasizes transition instants and a windowed spectral entropy $H(t)$ that captures changes in spectral structure.
These auxiliary features, used alongside the raw magnetic-field time series, aided event localization and peak-amplitude extraction across distances without requiring a detailed source model.

Using the event-resolved amplitudes, we quantified distance-dependent trends on log--log scales.
The door-related signature exhibited a dipole-like decay approaching $\Delta B_{\mathrm{Door}}\propto D^{-3}$ over the measured range, consistent with an effectively localized disturbance near the entrance region.
In contrast, the car and counterweight signatures displayed distance-dependent effective slopes and curvature on log--log plots, indicating that a single universal power-law exponent is insufficient to describe practical elevator sources under a door-referenced standoff definition.
Together, these results provide an empirical benchmark for range-resolved magnetic-perturbation sensing in unshielded and realistic building environments using a scalar OPAM.

Future work will extend this study in several directions.
First, hardware and procedural upgrades are expected to reduce the directly measured in-band ASD toward the intrinsic limit of approximately 21~pT/$\sqrt{\mathrm{Hz}}$ estimated in Section~\ref{sec:hw} and to extend the closed-loop in-band response beyond 30~Hz. These upgrades include replacing the commercial bench-top bias supply with a low-noise current source, degaussing the chamber prior to characterization, increasing the DAQ resolution, and migrating the digital lock-in to a COTS FPGA-based implementation.
Second, controlled multi-sensor measurements (e.g., gradiometric configurations combining atomic and fluxgate sensors) and improved geometric referencing will help disentangle distributed contributors and reduce ambiguity in standoff definitions for complex sources.
Third, the present timing-marker framework can be combined with model-based or data-driven detectors (including OBF/MED-inspired feature fusion and machine-learning classification) to improve low-SNR event detection, tracking, and source characterization in operational magnetic anomaly and infrastructure monitoring applications.

\begin{acknowledgments}
We gratefully acknowledge Dr.\ Sangkyung Lee and Dr.\ Sin Hyuk Yim (Agency for Defense Development) for their invaluable comments, guidance, and constructive suggestions, which significantly improved this manuscript.
We also thank Jeongwoo Lee (OAQ; KAIST) for assistance with the literature search and helpful discussions.
\end{acknowledgments}

\appendix

\section{Measured data at various standoff distances}\label{sec:appendix}

\begin{figure}[!htbp]
    \centering
    \includegraphics[width=1.0\linewidth]{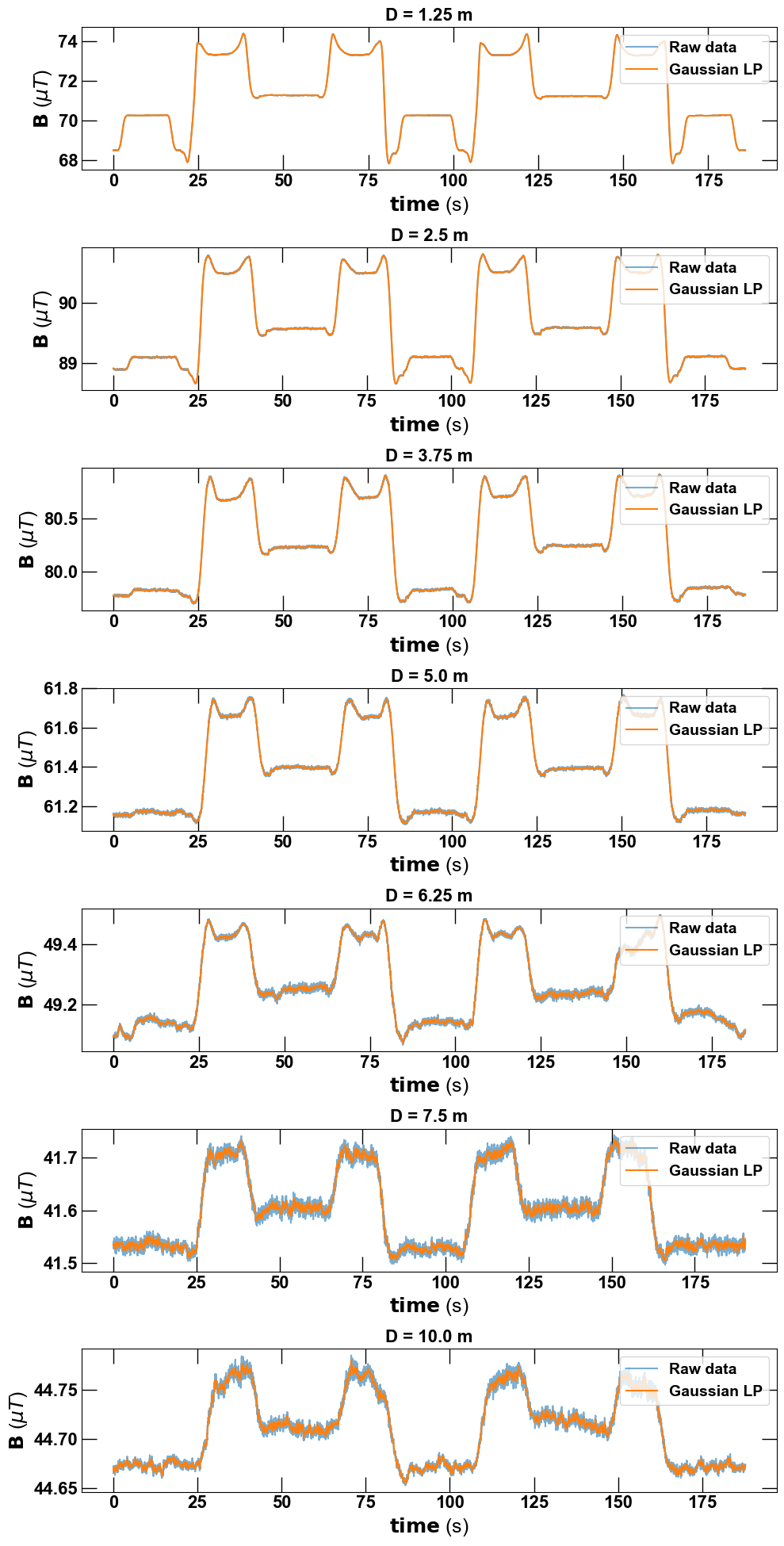}
    \caption{\label{fig:figapp1}
    Representative magnetic-field time traces measured at various standoff distances between the sensor and the elevator. The raw data are shown in blue, and the corresponding Gaussian-filtered data are shown in orange.}
\end{figure}

Figure~\ref{fig:figapp1} shows representative magnetic-field time traces measured at multiple standoff distances. For distances below approximately 5 m, magnetic-field variations associated with the opening and closing of the elevator doors are clearly observed in the time-domain data. At larger distances, the magnetic-field signatures generated by the door operation are no longer clearly resolved.

In contrast, magnetic-field perturbations associated with the motion of the elevator car and the counterweight system remain observable over the full range of measured distances, including the maximum standoff distance of 10 m. As the distance between the sensor and the elevator increases, the amplitude of the measured magnetic-field perturbations decreases systematically, reaching values on the order of 0.1 $\mu T$ at a distance of 10 m.

\newpage
\bibliography{aipsamp}% Produces the bibliography via BibTeX.

@article{degen_quantum_2017,
	title = {Quantum sensing},
	volume = {89},
	url = {https://link.aps.org/doi/10.1103/RevModPhys.89.035002},
	doi = {10.1103/RevModPhys.89.035002},
	number = {3},
	urldate = {2025-12-28},
	journal = {Reviews of Modern Physics},
	author = {Degen, C. L. and Reinhard, F. and Cappellaro, P.},
	month = jul,
	year = {2017},
	note = {Publisher: American Physical Society},
	pages = {035002},
}

@article{budker_optical_2007,
	title = {Optical magnetometry},
	volume = {3},
	copyright = {2007 Springer Nature Limited},
	issn = {1745-2481},
	url = {https://www.nature.com/articles/nphys566},
	doi = {10.1038/nphys566},
	number = {4},
	urldate = {2025-12-28},
	journal = {Nature Physics},
	author = {Budker, Dmitry and Romalis, Michael},
	month = apr,
	year = {2007},
	note = {Publisher: Nature Publishing Group},
	pages = {227--234},
}

@article{yim_note_2018,
	title = {Note: {Double}-layered polyimide film heater with low magnetic field generation},
	volume = {89},
	issn = {0034-6748},
	shorttitle = {Note},
	url = {https://doi.org/10.1063/1.5040398},
	doi = {10.1063/1.5040398},
	number = {11},
	urldate = {2025-12-28},
	journal = {Review of Scientific Instruments},
	author = {Yim, S. H. and Kim, Z. and Lee, S. and Kim, T. H. and Shim, K. M.},
	month = nov,
	year = {2018},
	pages = {116102},
}

@article{aslam_quantum_2023,
	title = {Quantum sensors for biomedical applications},
	volume = {5},
	copyright = {2023 Springer Nature Limited},
	issn = {2522-5820},
	url = {https://www.nature.com/articles/s42254-023-00558-3},
	doi = {10.1038/s42254-023-00558-3},
	number = {3},
	urldate = {2025-12-28},
	journal = {Nature Reviews Physics},
	author = {Aslam, Nabeel and Zhou, Hengyun and Urbach, Elana K. and Turner, Matthew J. and Walsworth, Ronald L. and Lukin, Mikhail D. and Park, Hongkun},
	month = mar,
	year = {2023},
	note = {Publisher: Nature Publishing Group},
	pages = {157--169},
}

@misc{muradoglu_quantum-assured_2025,
	title = {Quantum-assured magnetic navigation achieves positioning accuracy better than a strategic-grade {INS} in airborne and ground-based field trials},
	url = {http://arxiv.org/abs/2504.08167},
	doi = {10.48550/arXiv.2504.08167},
	urldate = {2025-12-28},
	publisher = {arXiv},
	author = {Muradoglu, Murat and Johnsson, Mattias T. and Wilson, Nathanial M. and Cohen, Yuval and Shin, Dongki and Navickas, Tomas and Pyragius, Tadas and Thomas, Divya and Thompson, Daniel and Moore, Steven I. and Rahman, Md Tanvir and Walker, Adrian and Dutta, Indranil and Bijjahalli, Suraj and Berlocher, Jacob and Hush, Michael R. and Anderson, Russell P. and Szigeti, Stuart S. and Biercuk, Michael J.},
	month = apr,
	year = {2025},
	note = {arXiv:2504.08167 [quant-ph]},
}

@article{canciani_absolute_2016,
	title = {Absolute {Positioning} {Using} the {{Earth}}'s {Magnetic} {Anomaly} {Field}},
	volume = {63},
	copyright = {Published 2016. This article is a U.S. Government work and is in the public domain in the USA},
	issn = {2161-4296},
	url = {https://onlinelibrary.wiley.com/doi/abs/10.1002/navi.138},
	doi = {10.1002/navi.138},
	number = {2},
	urldate = {2025-12-28},
	journal = {NAVIGATION},
	author = {Canciani, Aaron and Raquet, John},
	year = {2016},
	note = {\_eprint: https://onlinelibrary.wiley.com/doi/pdf/10.1002/navi.138},
	pages = {111--126},
}

@article{kotowski_semi-airborne_2025,
	title = {Semi-airborne electromagnetic exploration of deep sulfide deposits with {UAV}-towed magnetometers — {Part} 1: {Processing} and modeling},
	volume = {90},
	issn = {0016-8033},
	shorttitle = {Semi-airborne electromagnetic exploration of deep sulfide deposits with {UAV}-towed magnetometers — {Part} 1},
	url = {https://doi.org/10.1190/geo2024-0453.1},
	doi = {10.1190/geo2024-0453.1},
	number = {3},
	urldate = {2025-12-28},
	journal = {Geophysics},
	author = {Kotowski, Philipp O. and Becken, Michael and Rochlitz, Raphael and Schmalzl, Jörg and Ueding, Stefan and Tolksdorf, Pia and Wilhelm, Alex and Symons, Gregory},
	month = apr,
	year = {2025},
	pages = {WA261--WA274},
}

@article{sharma_magnetic_1987,
	title = {Magnetic method applied to mineral exploration},
	volume = {2},
	issn = {0169-1368},
	url = {https://www.sciencedirect.com/science/article/pii/0169136887900102},
	doi = {10.1016/0169-1368(87)90010-2},
	number = {4},
	urldate = {2025-12-28},
	journal = {Ore Geology Reviews},
	author = {Sharma, P. V.},
	month = aug,
	year = {1987},
	pages = {323--357},
}

@article{koss_optically_2022,
	title = {Optically {Pumped} {Magnetometer} {Measuring} {Fatigue}-{Induced} {Damage} in {Steel}},
	volume = {12},
	copyright = {http://creativecommons.org/licenses/by/3.0/},
	issn = {2076-3417},
	url = {https://www.mdpi.com/2076-3417/12/3/1329},
	doi = {10.3390/app12031329},
	number = {3},
	urldate = {2025-12-28},
	journal = {Applied Sciences},
	author = {Koss, Peter A. and Durmaz, Ali Riza and Blug, Andreas and Laskin, Gennadii and Pawar, Omkar Satish and Thiemann, Kerstin and Bertz, Alexander and Straub, Thomas and Elsässer, Christian},
	month = jan,
	year = {2022},
	note = {Publisher: Multidisciplinary Digital Publishing Institute},
	pages = {1329},
}

@article{huang_mains_2024,
	title = {{MAINS}: {A} {Magnetic}-{Field}-{Aided} {Inertial} {Navigation} {System} for {Indoor} {Positioning}},
	volume = {24},
	issn = {1558-1748},
	shorttitle = {{MAINS}},
	url = {https://ieeexplore.ieee.org/document/10480297},
	doi = {10.1109/JSEN.2024.3379932},
	number = {9},
	urldate = {2025-12-28},
	journal = {IEEE Sensors Journal},
	author = {Huang, Chuan and Hendeby, Gustaf and Fourati, Hassen and Prieur, Christophe and Skog, Isaac},
	month = may,
	year = {2024},
	pages = {15156--15166},
}

@article{sander-thommes_active_2023,
	title = {Active field compensation using optically pumped magnetometers},
	volume = {2},
	copyright = {Copyright (c) 2023 Proceedings on Automation in Medical Engineering},
	url = {https://www.journals.infinite-science.de/index.php/automed/article/view/772},
	number = {1},
	urldate = {2025-12-28},
	journal = {Proceedings on Automation in Medical Engineering},
	author = {Tilmann Sander-Thömmes and Yoshiaki Adachi},
	month = mar,
	year = {2023},
	pages = {772--772},
}

@article{rushton_unshielded_2022,
	title = {Unshielded portable optically pumped magnetometer for the remote detection of conductive objects using eddy current measurements},
	volume = {93},
	issn = {0034-6748},
	url = {https://doi.org/10.1063/5.0102402},
	doi = {10.1063/5.0102402},
	number = {12},
	urldate = {2025-12-28},
	journal = {Review of Scientific Instruments},
	author = {Rushton, L. M. and Pyragius, T. and Meraki, A. and Elson, L. and Jensen, K.},
	month = dec,
	year = {2022},
	pages = {125103},
}

@article{dang_ultrahigh_2010,
	title = {Ultrahigh sensitivity magnetic field and magnetization measurements with an atomic magnetometer},
	volume = {97},
	issn = {0003-6951},
	url = {https://doi.org/10.1063/1.3491215},
	doi = {10.1063/1.3491215},
	number = {15},
	urldate = {2025-12-28},
	journal = {Applied Physics Letters},
	author = {Dang, H. B. and Maloof, A. C. and Romalis, M. V.},
	month = oct,
	year = {2010},
	pages = {151110},
}

@article{oelsner_integrated_2022,
	title = {Integrated {Optically} {Pumped} {Magnetometer} for {Measurements} within {{Earth}}’s {Magnetic} {Field}},
	volume = {17},
	doi = {10.1103/PhysRevApplied.17.024034},
	number = {2},
	journal = {Physical Review Applied},
	author = {Oelsner, G.},
	year = {2022},
}

@article{Grujic2013,
    author  = {Gruji\'c, Z. D. and Weis, A.},
    title   = {Atomic magnetic resonance induced by amplitude-, frequency-, or polarization-modulated light},
    journal = {Phys. Rev. A},
    volume  = {88},
    pages   = {012508},
    year    = {2013},
    doi     = {10.1103/PhysRevA.88.012508}
}

@article{Tsyganok2019,
    author        = {Tsyganok, V. V. and Pershin, D. A. and Davletov, E. T. and Khlebnikov, V. A. and Akimov, A. V.},
    title         = {Scalar, tensor, and vector polarizability of {{Tm}} atoms in a 532-nm dipole trap},
    journal       = {Phys. Rev. A},
    volume        = {100},
    number        = {4},
    pages         = {042502},
    year          = {2019},
    doi           = {10.1103/PhysRevA.100.042502},
    archivePrefix = {arXiv},
    eprint        = {1905.03842},
    primaryClass  = {physics.atom-ph}
}

@article{yalaz_response_2024,
	title = {Response of {Lock}-{In} {Detection}-{Based} {Optically} {Pumped} {Magnetometers} to {High} {Bandwidth} {Magnetic} {Signals}},
	volume = {8},
	issn = {2475-1472},
	url = {https://ieeexplore.ieee.org/abstract/document/10598236},
	doi = {10.1109/LSENS.2024.3427435},
	number = {8},
	urldate = {2025-12-28},
	journal = {IEEE Sensors Letters},
	author = {Yalaz, Mevlüt and Elzenheimer, Eric and Wolframm, Henrik and Knappe-Grüneberg, Silvia and Höft, Michael and Deuschl, Günther and Helmers, Ann-Kristin and Sander, Tilmann},
	month = aug,
	year = {2024},
	pages = {1--4},
}

@article{schultze_optically_2017,
	title = {An {Optically} {Pumped} {Magnetometer} {Working} in the {Light}-{Shift} {Dispersed} {Mz} {Mode}},
	volume = {17},
	copyright = {http://creativecommons.org/licenses/by/3.0/},
	issn = {1424-8220},
	url = {https://www.mdpi.com/1424-8220/17/3/561},
	doi = {10.3390/s17030561},
	number = {3},
	urldate = {2025-12-28},
	journal = {Sensors},
	author = {Schultze, Volkmar and Schillig, Bastian and IJsselsteijn, Rob and Scholtes, Theo and Woetzel, Stefan and Stolz, Ronny},
	month = mar,
	year = {2017},
	note = {Publisher: Multidisciplinary Digital Publishing Institute},
	pages = {561},
}

@article{bell_optically_1961,
	title = {Optically {Driven} {Spin} {Precession}},
	volume = {6},
	url = {https://link.aps.org/doi/10.1103/PhysRevLett.6.280},
	doi = {10.1103/PhysRevLett.6.280},
	number = {6},
	urldate = {2025-12-28},
	journal = {Physical Review Letters},
	author = {Bell, William E. and Bloom, Arnold L.},
	month = mar,
	year = {1961},
	note = {Publisher: American Physical Society},
	pages = {280--281},
}

@misc{yoon_laser_2025,
	title = {Laser mode-hopping assisted all-optical single beam pulsed atomic magnetometer},
	url = {http://arxiv.org/abs/2404.01874},
	doi = {10.48550/arXiv.2404.01874},
	urldate = {2025-12-28},
	publisher = {arXiv},
	author = {Yoon, Ji Hoon and Hong, Sang Hyuk and Jeong, Taek and Yim, Sin Hyuk and Shim, Kyu Min and Lee, Sangkyung},
	month = jan,
	year = {2025},
	note = {arXiv:2404.01874 [physics]},
}

@article{VanBaak1996,
    author  = {Van Baak, D. A.},
    title   = {Resonant {{Faraday}} rotation as a probe of atomic dispersion},
    journal = {American Journal of Physics},
    volume  = {64},
    number  = {6},
    pages   = {724--735},
    year    = {1996},
    doi     = {10.1119/1.18356}
}

@article{petrenko_single-beam_2021,
	title = {Single-{Beam} {All}-{Optical} {Nonzero}-{Field} {Magnetometric} {Sensor} for {Magnetoencephalography} {Applications}},
	volume = {15},
	url = {https://link.aps.org/doi/10.1103/PhysRevApplied.15.064072},
	doi = {10.1103/PhysRevApplied.15.064072},
	number = {6},
	urldate = {2025-12-28},
	journal = {Physical Review Applied},
	author = {Petrenko, M.V. and Pazgalev, A.S. and Vershovskii, A.K.},
	month = jun,
	year = {2021},
	note = {Publisher: American Physical Society},
	pages = {064072},
}

@article{zhang_zero_2022,
	title = {Zero field optically pumped magnetometer with independent dual-mode operation},
	volume = {20},
	copyright = {© 2022 Chinese Laser Press},
	url = {https://opg.optica.org/col/abstract.cfm?uri=col-20-8-081202},
	number = {8},
	urldate = {2025-12-28},
	journal = {Chinese Optics Letters},
	author = {Zhang, Shaowen and Lu, Jixi and Zhou, Ying and Lu, Fei and Yin, Kaifeng and Zhan, Di and Zhai, Yueyang and Ye, Mao},
	month = aug,
	year = {2022},
	note = {Publisher: Chinese Optical Society},
	pages = {081202},
}

@article{wu_compact_2025,
	title = {Compact high-bandwidth single-beam optically-pumped magnetometer for biomagnetic measurement},
	volume = {16},
	copyright = {© 2024 Optica Publishing Group},
	issn = {2156-7085},
	url = {https://opg.optica.org/boe/abstract.cfm?uri=boe-16-1-235},
	doi = {10.1364/BOE.545624},
	number = {1},
	urldate = {2025-12-28},
	journal = {Biomedical Optics Express},
	author = {Wu, Tianbo and Xiao, Wei and Peng, Xiang and Wu, Teng and Guo, Hong},
	month = jan,
	year = {2025},
	note = {Publisher: Optica Publishing Group},
	pages = {235--244},
}

@article{imajo_signal_2021,
	title = {Signal and {Noise} {Separation} {From} {Satellite} {Magnetic} {Field} {Data} {Through} {Independent} {Component} {Analysis}: {Prospect} of {Magnetic} {Measurements} {Without} {Boom} and {Noise} {Source} {Information}},
	volume = {126},
	copyright = {© 2021. American Geophysical Union. All Rights Reserved.},
	issn = {2169-9402},
	shorttitle = {Signal and {Noise} {Separation} {From} {Satellite} {Magnetic} {Field} {Data} {Through} {Independent} {Component} {Analysis}},
	url = {https://onlinelibrary.wiley.com/doi/abs/10.1029/2020JA028790},
	doi = {10.1029/2020JA028790},
	number = {5},
	urldate = {2025-12-28},
	journal = {Journal of Geophysical Research: Space Physics},
	author = {Imajo, S. and Nosé, M. and Aida, M. and Matsumoto, H. and Higashio, N. and Tokunaga, T. and Matsuoka, A.},
	year = {2021},
	note = {\_eprint: https://agupubs.onlinelibrary.wiley.com/doi/pdf/10.1029/2020JA028790},
	pages = {e2020JA028790},
}

@article{BudkerYashchukZolotorev1998,
  author  = {Budker, Dmitry and Yashchuk, Valeriy and Zolotorev, Max},
  title   = {Nonlinear Magneto-optic Effects with Ultranarrow Widths},
  journal = {Physical Review Letters},
  volume  = {81},
  number  = {26},
  pages   = {5788--5791},
  year    = {1998},
  doi     = {10.1103/PhysRevLett.81.5788}
}

@article{Yoo2023copp,
    author = {Jae-Keun Yoo and Deok-Young Lee and Sin Hyuk Yim and Hyun-Gue Hong and Sun Do Lim and Seung Kwan Kim and Young-Pyo Hong and No-Weon Kang and In-Ho Bae},
    journal = {Curr. Opt. Photon.},
    number = {2},
    pages = {207--212},
    publisher = {Optica Publishing Group},
    title = {Fabrication of High-purity {Rb} Vapor Cell for Electric Field Sensing},
    volume = {7},
    month = {Apr},
    year = {2023},
    url = {https://opg.optica.org/copp/abstract.cfm?URI=copp-7-2-207},
}

@article{Yim2022AIP,
    author = {Yim, S. H. and Lee, D.-Y. and Lee, S. and Kim, M. M.},
    title = {Experimental setup to fabricate {Rb}–{Xe} gas cells for atom spin gyroscopes},
    journal = {AIP Advances},
    volume = {12},
    number = {1},
    pages = {015025},
    year = {2022},
    month = {01},
    issn = {2158-3226},
    doi = {10.1063/5.0069211},
    url = {https://doi.org/10.1063/5.0069211},
}

@article{Lee_cubic,
    author  = {Lee, D.-Y. and Lee, S. and Kim, M. M. and Yim, S. H.},
    title   = {Magnetic-field-inhomogeneity-induced transverse-spin relaxation of gaseous $^{129}${Xe} in a cubic cell with a stem},
    journal = {Phys. Rev. A},
    volume  = {104},
    number  = {4},
    pages   = {042819},
    year    = {2021},
    doi     = {10.1103/PhysRevA.104.042819},
    url     = {https://doi.org/10.1103/PhysRevA.104.042819},
}

@article{zanoni_picotesla_2024,
    author  = {Zanoni, A. and Mouloudakis, K. and Tayler, M. C. D. and Corrielli, G. and Osellame, R. and Mitchell, M. W. and Lucivero, V. G.},
    title   = {Picotesla optically pumped magnetometer using a laser-written vapor cell with sub-mm cross section},
    journal = {J. Appl. Phys.},
    volume  = {136},
    number  = {14},
    pages   = {144401},
    year    = {2024},
    doi     = {10.1063/5.0230180},
    url     = {https://doi.org/10.1063/5.0230180},
}

@article{Kim2023JAP,
    author = {Kim, M. M. and Lee, S. and Yim, S. H. and Yoon, J. H.},
    title = {A low-magnetic packaging for a distributed Bragg reflector laser diode chip for atomic sensor applications},
    journal = {Journal of Applied Physics},
    volume = {133},
    number = {16},
    pages = {164502},
    year = {2023},
    month = {04},
    issn = {0021-8979},
    doi = {10.1063/5.0141434},
    url = {https://doi.org/10.1063/5.0141434},
}

@article{Stolz2022MinerEcon,
    author  = {Stolz, R. and Schiffler, M. and Becken, M.},
    title   = {{SQUIDs} for magnetic and electromagnetic methods in mineral exploration},
    journal = {Mineral Economics},
    volume  = {35},
    number  = {},
    pages   = {467--494},
    year    = {2022},
    month   = {},
    issn    = {},
    doi     = {10.1007/s13563-022-00333-3},
    url     = {https://doi.org/10.1007/s13563-022-00333-3},
    eprint  = {}
}

@article{Yao2022Opt,
    author = {Han Yao and Benjamin Maddox and F. Renzoni},
    journal = {Opt. Express},
    number = {23},
    pages = {42015--42025},
    publisher = {Optica Publishing Group},
    title = {High-sensitivity operation of an unshielded single cell radio-frequency atomic magnetometer},
    volume = {30},
    month = {Nov},
    year = {2022},
    url = {https://opg.optica.org/oe/abstract.cfm?URI=oe-30-23-42015},
    doi = {10.1364/OE.476016},
}

@article{Kang2017JournalSensors,
  author    = {Kang, Chong and Fan, Liming and Zheng, Quan and Kang, Xiyuan and Zhou, Jian and Zhang, Xiaojun},
  title     = {Experimental Study on the Localization of Moving Object by Total Geomagnetic Field},
  journal   = {Journal of Sensors},
  volume    = {2017},
  pages     = {Article ID 7948930},
  year      = {2017},
  month     = {07},
  doi       = {10.1155/2017/7948930},
  url       = {https://doi.org/10.1155/2017/7948930},
}

@Article{rs16020363,
AUTHOR = {Liu, Xingen and Yuan, Zifan and Du, Changping and Peng, Xiang and Guo, Hong and Xia, Mingyao},
TITLE = {Adaptive Basis Function Method for the Detection of an Undersurface Magnetic Anomaly Target},
JOURNAL = {Remote Sensing},
VOLUME = {16},
YEAR = {2024},
NUMBER = {2},
ARTICLE-NUMBER = {363},
URL = {https://www.mdpi.com/2072-4292/16/2/363},
ISSN = {2072-4292},
DOI = {10.3390/rs16020363}
}

@article{Fan2020JoS,
author = {Fan, Liming and Kang, Chong and Wang, Huigang and Hu, Hao and Zou, Mingliang},
title = {Adaptive Magnetic Anomaly Detection Method with Ensemble Empirical Mode Decomposition and Minimum Entropy Feature},
journal = {Journal of Sensors},
volume = {2020},
number = {1},
pages = {8856577},
doi = {https://doi.org/10.1155/2020/8856577},
url = {https://onlinelibrary.wiley.com/doi/abs/10.1155/2020/8856577},
eprint = {https://onlinelibrary.wiley.com/doi/pdf/10.1155/2020/8856577},
year = {2020}
}

@article{Wang2021AIP,
    author = {Wang, Z. and Qiu, J. and Xie, D. and Ou, J. and Xu, Q.},
    title = {Weak magnetic anomaly signal detection based on the entropy of mixed differential signal},
    journal = {AIP Advances},
    volume = {11},
    number = {1},
    pages = {015013},
    year = {2021},
    month = {01},
    issn = {2158-3226},
    doi = {10.1063/9.0000080},
    url = {https://doi.org/10.1063/9.0000080},
}

@article{ZHOU2023,
title = {Magnetic anomaly detection via a combination approach of minimum entropy and gradient orthogonal functions},
journal = {ISA Transactions},
volume = {134},
pages = {548-560},
year = {2023},
issn = {0019-0578},
doi = {https://doi.org/10.1016/j.isatra.2022.08.026},
url = {https://www.sciencedirect.com/science/article/pii/S0019057822004293},
author = {Jiaqi Zhou and Chengdong Wang and Genzhai Peng and Huan Yan and Zhihong Zhang and Yong Chen}
}

@inproceedings{Martinez2020ISMRM,
  title     = {{MRI} Room Placement: The effect of an elevator operation on local magnetic environment},
  author    = {Martinez, Diego F. and Davieau, Kieffer J. and Lessard, Eric J. and Handler, William B. and Chronik, Blaine A.},
  booktitle = {Proceedings of the International Society for Magnetic Resonance in Medicine (ISMRM) Annual Meeting},
  year      = {2020},
}

@article{CHEN2022113806,
title = {Modeling and experimental investigation of magnetic anomaly detection using advanced triaxial magnetoelectric sensors},
journal = {Sensors and Actuators A: Physical},
volume = {346},
pages = {113806},
year = {2022},
issn = {0924-4247},
doi = {https://doi.org/10.1016/j.sna.2022.113806},
url = {https://www.sciencedirect.com/science/article/pii/S0924424722004411},
author = {Ziyun Chen and Wenning Di and Rui Chen and Tingyu Deng and Yuhang Wang and Haoran You and Li Lu and Tao Han and Jie Jiao and Haosu Luo},
}

@Article{s24124028,
AUTHOR = {Li, Hangcheng and Luo, Jiaming and Zhang, Jiajun and Li, Jing and Zhang, Yi and Zhang, Wenwei and Zhang, Mingji},
TITLE = {Determinants of Maximum Magnetic Anomaly Detection Distance},
JOURNAL = {Sensors},
VOLUME = {24},
YEAR = {2024},
NUMBER = {12},
ARTICLE-NUMBER = {4028},
URL = {https://www.mdpi.com/1424-8220/24/12/4028},
PubMedID = {38931811},
ISSN = {1424-8220},DOI = {10.3390/s24124028}
}

@article{Mehta2025APL_deadzone,
  author    = {Mehta, Shrey and Samanta, G. K. and Grewal, Raghwinder Singh},
  title     = {Dead-zone-free single-beam atomic magnetometer based on free-induction-decay of {Rb} atoms},
  journal   = {Applied Physics Letters},
  volume    = {126},
  number    = {4},
  pages     = {044002},
  year      = {2025},
  doi       = {10.1063/5.0248330},
}

@article{Jiang2025OL_FIDenhance,
  author    = {Jiang, Liwei and Xu, Jinghong and Liu, Junhao and Fang, Chi and Zhu, Jun and Zou, Yuntian and Chen, Yuanqiang and Liu, Jiali and Zhao, Xin},
  title     = {Signal enhancement in single-beam free-induction-decay magnetometers by manipulation of spin polarization},
  journal   = {Optics Letters},
  volume    = {50},
  number    = {12},
  pages     = {4014--4017},
  year      = {2025},
  doi       = {10.1364/OL.558171},
}

@article{IEEE_TIM2019_singlebeam,
  author={Levy, Cooper S. and Kornack, Thomas W. and Mercier, Patrick P.},
  journal={IEEE Transactions on Instrumentation and Measurement},
  title={{{Bell–Bloom}} Magnetometer Linearization by Intensity Modulation Cancellation},
  year={2020},
  volume={69},
  number={3},
  pages={883-892},
  doi={10.1109/TIM.2019.2904373}
  }

@article{aleksandrov_modern_2009,
  author  = {Aleksandrov, E. B. and Vershovskii, A. K.},
  title   = {Modern radio-optical methods in quantum magnetometry},
  journal = {Physics-Uspekhi},
  volume  = {52},
  number  = {6},
  pages   = {573--601},
  year    = {2009},
  doi     = {10.3367/UFNe.0179.200906f.0605}
}

@article{schwindt_chip-scale_2007,
  author  = {Schwindt, Peter D. D. and Lindseth, Brad and Knappe, Svenja and Shah, Vishal and Kitching, John and Liew, Li-Anne},
  title   = {Chip-scale atomic magnetometer with improved sensitivity by use of the {$M_x$} technique},
  journal = {Applied Physics Letters},
  volume  = {90},
  number  = {8},
  pages   = {081102},
  year    = {2007},
  doi     = {10.1063/1.2709532}
}

\end{document}